\documentclass[aps,prd,twocolumn,nofootinbib,superscriptaddress,longbibliography,amsfonts,amssymb,amsmath,titlepage]{revtex4-1}

\usepackage[usenames,dvipsnames]{color}
\usepackage{microtype}
\usepackage{IEEEtrantools}
\usepackage{mathrsfs}
\usepackage{amsthm}
\usepackage{enumitem}
\usepackage[normalem]{ulem}
\usepackage{graphicx}
\usepackage{stackengine}
\usepackage[caption=false]{subfig}

\usepackage[pdftex]{hyperref}

\begin{document}

\title{Gravitational-wave parameter estimation with autoregressive
  neural network flows}

\author{Stephen R. Green}
\email{stephen.green@aei.mpg.de}
\affiliation{Max Planck Institute for Gravitational Physics (Albert Einstein Institute)\\
  Am M\"uhlenberg 1, 14476 Potsdam, Germany}

\author{Christine Simpson}
\email{christine.simpson@ed.ac.uk}
\affiliation{School of Informatics, University of Edinburgh, 10 Crichton Street, Edinburgh,
EH8 9AB, United Kingdom}

\author{Jonathan Gair}
\email{jonathan.gair@aei.mpg.de}
\affiliation{Max Planck Institute for Gravitational Physics (Albert Einstein Institute)\\
  Am M\"uhlenberg 1, 14476 Potsdam, Germany}

\begin{abstract}
  We introduce the use of autoregressive normalizing flows for rapid
  likelihood-free inference of binary black hole system parameters
  from gravitational-wave data with deep neural networks. A
  normalizing flow is an invertible mapping on a sample space that can
  be used to induce a transformation from a simple probability
  distribution to a more complex one: if the simple distribution can
  be rapidly sampled and its density evaluated, then so can the
  complex distribution.  Our first application to gravitational waves
  uses an autoregressive flow, conditioned on detector strain data, to
  map a multivariate standard normal distribution into the posterior
  distribution over system parameters.  We train the model on
  artificial strain data consisting of \verb+IMRPhenomPv2+ waveforms
  drawn from a five-parameter $(m_1, m_2, \phi_0, t_c, d_L)$ prior and
  stationary Gaussian noise realizations with a fixed power spectral
  density. This gives performance comparable to current best
  deep-learning approaches to gravitational-wave parameter
  estimation. We then build a more powerful latent variable model by
  incorporating autoregressive flows within the variational
  autoencoder framework. This model has performance comparable to
  Markov chain Monte Carlo and, in particular, successfully models the
  multimodal $\phi_0$ posterior. Finally, we train the autoregressive
  latent variable model on an expanded parameter space, including also
  aligned spins $(\chi_{1z}, \chi_{2z})$ and binary inclination
  $\theta_{JN}$, and show that all parameters and degeneracies are
  well-recovered. In all cases, sampling is extremely fast, requiring
  less than two seconds to draw $10^4$ posterior samples.
\end{abstract}

\maketitle

\section{Introduction}

The task of gravitational-wave parameter estimation is to determine
the system parameters that gave rise to observed detector strain
data. This is accomplished using Bayesian inference. Assuming a
\emph{likelihood} model $p(y|x)$ for strain data $y$ conditioned on
system parameters $x$, and a \emph{prior} distribution $p(x)$, one
obtains through Bayes' theorem the \emph{posterior} distribution,
\begin{equation}
  \label{eq:bayes}
  p(x|y) = \frac{p(y|x)p(x)}{p(y)},
\end{equation}
where the normalization $p(y)$ is called the model
\emph{evidence}. Generally, one can evaluate the likelihood explicitly
(although this may be computationally expensive) and the prior is also
known, so this allows for calculation of the posterior up to
normalization. One can then use an algorithm such as Markov chain
Monte Carlo (MCMC) to obtain samples from the posterior.

Standard sampling algorithms are effective, but they can be
computationally costly. Indeed, for binary black holes, obtaining a
sufficient number of posterior samples can take on the order of days,
whereas for binary neutron stars, this extends to weeks. Especially
for binary neutron stars, which might have multimessenger
counterparts, it is critical to reduce this time to provide accurate
information to electromagnetic observers.

There have recently been several efforts to speed up parameter
estimation by using deep
learning~\cite{Chua:2019wwt,Gabbard:2019rde,Shen:2019vep}. The main
tool of deep learning is the neural network, which is a trainable and
very flexible function approximator. Neural networks can have millions
of parameters, which are tuned through stochastic gradient descent to
optimize a loss function. In this way, very complex functions can be
approximated. Since conditional distributions can be parametrized by
functions, neural networks can be used to model gravitational-wave
posteriors.

A key observation of~\cite{Chua:2019wwt,Gabbard:2019rde} is that
Bayes' rule can be applied in such a way that training only requires
samples from the prior and the gravitational-wave likelihood, i.e.,
strain realizations $y \sim p(y|x)$. Although the network learns the
posterior distribution, posterior samples are not needed for
training. Training also does not require any likelihood evaluations,
and for this reason this approach is known as \emph{likelihood-free
  inference}. Since obtaining posterior samples and evaluating the
likelihood are expensive, the likelihood-free approach is very fast in
comparison. It is also applicable in contexts where a simulation of
the data generative process is available, but the likelihood function
may be unknown.

In~\cite{Chua:2019wwt}, this approach was applied successfully with a
multivariate Gaussian posterior model, i.e.,
$p(x|y) = \mathcal{N}(\mu(y), \Sigma(y))$, where the mean $\mu(y)$ and
covariance matrix $\Sigma(y)$ are given by neural networks. Once
trained on simulated waveforms and noise realizations, the Gaussian
posterior model is trivial to sample.

The challenge, then, is to define a sufficiently flexible model
distribution for the posterior. Gaussians are good approximations for
very high signal-to-noise ratio, but posteriors can in general have
higher moments and multimodality. Ref.~\cite{Chua:2019wwt} also
experimented with Gaussian mixture models and posterior histogram
outputs. The performance of the Gaussian mixture, however, was not a
significant improvement over the single Gaussian, and although
histogram was effective, it is limited to describing one or two
parameters.

A promising approach to increase the flexibility of a posterior model
is to introduce latent variables $z$ and perform variational Bayesian
inference. The posterior is then given by marginalization over the
latent variables, i.e., $p(x|y) = \int p(x|z,y)p(z|y)dz$. With
$p(x|z,y)$ and $p(z|y)$ both taken to be Gaussian, this nevertheless
gives \emph{non-}Gaussian $p(x|y)$. This can be implemented using the
variational autoencoder
framework~\cite{Kingma:2013,rezende2014stochastic}, and recent
results~\cite{Gabbard:2019rde} for gravitational-wave parameter
estimation with a conditional variational autoencoder (CVAE) have
demonstrated the recovery of non-Gaussian posteriors (although not
multimodality).

In this work, we use the method of \emph{normalizing
  flows}~\cite{rezende2015variational} (specifically, \emph{masked
  autoregressive
  flows}~\cite{kingma2016improved,papamakarios2017masked}) to increase
the flexibility of the gravitational-wave posterior. A normalizing
flow $f: X \to X$ is an invertible mapping on a sample space $X$, with
simple Jacobian determinant. Using the change of variables rule for
probabilities, this induces a transformation,
\begin{equation}
  \label{eq:changeofvariables}
  p(x) = \pi(f^{-1}(x))\left| \det \frac{\partial(f^{-1}_1,\ldots,f^{-1}_n)}{\partial( x_1,\ldots, x_n)} \right|,
\end{equation}
from a base distribution $\pi(u)$ to a new distribution $p(x)$. Here
$n=\dim X$. Starting from a simple standard normal distribution for
$\pi(u)$, one can obtain a more complex distribution by applying a
normalizing flow. To describe conditional distributions, such as
gravitational-wave posteriors, a normalizing flow may also be
conditioned on additional variables; for our application, we condition
$f$ on detector strain data $y$.

The fact that $f$ is invertible means that the normalizing flow
enables both sampling and density evaluation of $p(x)$, provided this
is possible for $\pi(u)$. To sample, one first samples
$u \sim \pi(u)$, and then $x = f(u)$ is a sample from $p(x)$. For
given $x$, to evaluate the density, the inverse mapping
$u = f^{-1}(x)$ is used in the right hand side
of~\eqref{eq:changeofvariables}.

Normalizing flows may be used to model gravitational-wave posterior
distributions directly, and this is the first application that we
describe. We fix the base distribution to be multivariate standard
normal of the same dimension as the system parameter space. We then
take the basic flow unit to be a three-hidden-layer MADE neural
network~\cite{germain2015made} conditioned on $y$, and we stack
several of these to obtain a sufficiently flexible posterior
model. This is called a masked autoregressive flow
(MAF)~\cite{papamakarios2017masked}. Since the density can be
evaluated directly through~\eqref{eq:changeofvariables}, the network
can be trained using stochastic gradient descent to maximize the
likelihood over the network parameters that the training data ($(x,y)$
pairs) came from the model.

Normalizing flows can also be incorporated into the CVAE: they can be
used to enhance the flexibility of the encoder, decoder, and prior
component networks, thereby increasing the overall flexibility of the
model~\cite{kingma2016improved,chen2016variational}.  In
section~\ref{sec:prelim} of this work, we describe all of the above
networks in further detail.

In section~\ref{sec:case1} we present the results of our experiments
in using these models to describe gravitational-wave posteriors over a
five-dimensional space of system
parameters. Following~\cite{Gabbard:2019rde}, we study binary black
hole waveforms over the parameters $(m_1, m_2, \phi_0, t_c, d_L)$,
with added noise drawn from a stationary Gaussian distribution with
fixed power spectral density. We work in the time domain. We find that
the basic MAF network achieves results comparable
to~\cite{Gabbard:2019rde}, but with the advantage of not having any
latent variables to marginalize over. In our experiments, however,
neither of these networks successully models the posterior over
$\phi_0$, which is multimodal. We next test the more powerful
autoregressive CVAE, which does succeed in modeling the multimodality
in $\phi_0$. To validate our recovered posteriors, we present p--p
plots consistent with uniformly distributed percentile scores in each
parameter, as well as comparisons to MCMC sampling.

We then expand the parameter space to include aligned spins
$(\chi_{1z}, \chi_{2z})$ and binary inclination $\theta_{JN}$ in
section~\ref{sec:spins}. We train the CVAE network with autoregressive
flows to model the posterior over this eight-dimensional space. We
find that the network once again successfully models all
parameters. This includes the various degeneracies, e.g., between
$\chi_{1z}$ and $\chi_{2z}$, and between $\theta_{JN}$ and $d_L$. We
validate our results again with a p--p plot.

This work is organized as follows. In the following section we
describe in more detail the various types of neural networks that we
use for parameter estimation. In section~\ref{sec:case1} we describe
our experiments with the five-dimensional parameter space, and in
section~\ref{sec:spins} the enlarged eight-dimensional space. Finally,
in section~\ref{sec:conclusions} we conclude with a discussion of
potential further improvements.

\emph{Software:} All of our neural networks are implemented in
PyTorch~\cite{NEURIPS2019_9015}, with the autoregressive network
implemented using Pyro~\cite{bingham2019pyro}. We used
emcee~\cite{Foreman_Mackey_2013} to generate MCMC comparisons, and
ChainConsumer~\cite{Hinton2016} to produce corner plots.

\emph{Notation:} The various vector spaces, along with their
dimensionalities are given in table~\ref{tab:spaces}.
\begin{table}[h]
  \caption{\label{tab:spaces}Vector spaces.}
  \begin{ruledtabular}
    \begin{tabular}{ccl}
      space & dimension & description \\
      \hline
      X & n & system parameters \\
      Y & m & strain data \\
      Z & l & latent variables
    \end{tabular}
  \end{ruledtabular}
\end{table}

\section{Preliminaries}\label{sec:prelim}

In this section we review important deep learning concepts, and we
discuss them in the context of gravitational-wave parameter
estimation. The first two subsections describe ideas that have already
been implemented for parameter estimation, in \cite{Chua:2019wwt} and
\cite{Gabbard:2019rde}, respectively, and the last two describe the
autoregressive flows that we explore in this work.

\subsection{Neural network models of gravitational-wave posteriors}
\label{sec:nn-gw}

Suppose we have a posterior distribution $p_{\text{true}}(x|y)$. Our
aim is to train a neural network to give an approximation $p(x|y)$ to
$p_{\text{true}}(x|y)$. The ``true'' posterior is itself a model for
the physical system. For gravitational waves, it is defined through
Bayes' theorem in terms of a prior $p_{\text{true}}(x)$ over the
system parameters $x$ and a likelihood $p_{\text{true}}(y|x)$. The
likelihood depends on a choice of waveform model $h$ and a measured
noise power spectral density $S_n(f)$: it is the probability density that
the residual $n = y - h(x)$ is drawn from a stationary Gaussian noise
distribution with PSD $S_n(f)$. For further details on the noise model
and likelihood, see, e.g., \cite{LIGOScientific:2019hgc}.

For stationary Gaussian noise and known $h(x)$, it is trivial to
sample from the likelihood. In contrast, the ``inverse problem'' of
sampling from the posterior---sampling over parameters $x$---requires
an algorithm such as MCMC and many evaluations of the waveform model
and the likelihood. This repeated comparison between model waveforms
and data is computationally expensive, and for this reason we wish to
develop the neural network model, $p(x|y)$.

The first step in developing the model is to parametrize the posterior
in terms of a neural network. For now, we take as our model a
multivariate normal distribution~\cite{Chua:2019wwt}, although our
main interest later will be in defining more flexible models. That is,
we take
\begin{align}\label{eq:gaussian}
  p(x|y) &= \frac{1}{\sqrt{(2\pi)^n|\det \Sigma(y)|}} \times \nonumber\\
         &\quad\exp\left( -\frac{1}{2}\sum_{ij=1}^n(x_i - \mu_i(y)) \Sigma^{-1}_{ij}(y) (x_j - \mu_j(y))  \right),
\end{align}
where the mean $\mu(y)$ and covariance matrix $\Sigma(y)$ are
functions of data $y$ defined by a feedforward neural network. (To
ensure $\Sigma(y)$ is positive definite and symmetric, it is in
practice better to take the Cholesky decomposition,
$\Sigma(y) = A(y)A(y)^\top$, where $A(y)$ is lower-triangular with
positive diagonal entries. $A(y)$ is then modeled with the neural
network.)

Feedforward neural networks can have a variety of configurations, but
the simplest consists of a sequence of fully-connected \emph{layers.}
The output of the first \emph{hidden} layer (of dimension $d_1$) is
\begin{equation}
  h_1 = \sigma( W_1 y + b_1 ),
\end{equation}
where $W_1$ is a $d_1 \times m$ matrix, and $b_1$ is a
$d_1$-dimensional vector. $\sigma$ is an element-wise nonlinearity,
typically taken to be a \emph{Rectified Linear Unit (ReLU)},
\begin{equation}
  \sigma(u) = \begin{cases}
    u & \text{if } u > 0,\\
    0 & \text{if } u \le 0.
  \end{cases}
\end{equation}
The output $h_1$ is then passed through a second hidden layer,
\begin{equation}
  h_2 = \sigma( W_2 h_1 + b_2 ),
\end{equation}
and so on, for as many hidden layers as desired. A final affine
transformation $W_ph_{p-1} + b_p$ is then applied,
and the outputs repackaged into a vector $\mu$ and lower-triangular
matrix $A$. A suitable nonlinearity should be applied to ensure
positive diagonal components of $A$, but otherwise the components of
$\mu$ and $A$ should be unconstrained. The weight matrices $W_i$ and
bias vectors $b_i$ are initialized randomly (see,
e.g.,~\cite{Goodfellow-et-al-2016}) and then trained through
stochastic gradient descent to optimize a loss function.

With the posterior defined, the loss function should be chosen so that
after training, $p(x|y)$ is a good approximation to
$p_{\text{true}}(x|y)$. We therefore take it to be the expectation
value (over $y$) of the cross-entropy between the two distributions,
\begin{equation}\label{eq:loss1}
  L = - \int dx dy \, p_{\text{true}}(x|y) p_{\text{true}}(y) \log p(x|y),
\end{equation}
i.e., we maximize the likelihood over the network parameters (the
weights and biases) that the training data came from the model.
Minimizing $L$ is equivalent to minimizing the expectation value of
the Kullback-Leibler (KL) divergence between the true and model
posteriors, since $p_{\text{true}}(x|y)$ is fixed.

The loss function in the form~\eqref{eq:loss1} is actually not ideal
for our purposes. The reason is that it requires sampling from
$p_{\text{true}}(x|y)$---a very costly operation. Instead, as pointed
out by~\cite{Chua:2019wwt,Gabbard:2019rde}, we can use Bayes' theorem
in a very advantageous way, to write
\begin{equation}\label{eq:Hloss}
  L = - \int dx dy \, p_{\text{true}}(y|x) p_{\text{true}}(x) \log p(x|y).
\end{equation}
To train the network now requires sampling from the likelihood, not
the posterior. On a minibatch of training data of size $N$, we
approximate
\begin{equation}
  L \approx - \frac{1}{N} \sum_{i=1}^N \log p(x^{(i)}|y^{(i)}),
\end{equation}
where $x^{(i)} \sim p_{\text{true}}(x)$ and
$y^{(i)} \sim p_{\text{true}}(y|x^{(i)})$. The loss function can be
explicitly evaluated using the expression~\eqref{eq:gaussian}.

The gradient of $L$ with respect to the network parameters can be
calculated using backpropagation (i.e., the chain rule), and the
network optimized with stochastic gradient descent. The training set
consists of parameter samples $x$ and waveforms $h(x)$; random noise
realizations are added at train time to obtain data samples $y$.

\subsection{Variational autoencoders}

One way to increase the flexibility of the model is to introduce
latent variables $z$, and define the gravitational-wave posterior
by first sampling from a prior over $z$, and then from a
distribution over $x$, conditional on $z$. In other words,
\begin{equation}\label{eq:posterior-latent}
  p(x|y) = \int dz \, p(x|z,y) p(z|y).
\end{equation}
If one takes $p(x|z,y)$ and $p(z|y)$ to both be multivariate Gaussians,
then $p(x|y)$ is a Gaussian mixture of Gaussians. In this way one can
describe a more flexible posterior.

At first glance it is not clear how to train such a model: the
posterior~\eqref{eq:posterior-latent} is intractable, since evaluation
involves marginalizing over $z$. If one knew the
posterior\footnote{The variational posterior $p(z|x,y)$ should not be
  confused with the gravitational-wave posterior $p(x|y)$. It should
  be clear from context to which posterior distribution we are
  referring.} $p(z|x,y)$, then
\begin{equation}
  p(x|y) = \frac{p(x|z,y)p(z|y)}{p(z|x,y)}
\end{equation}
could be evaluated directly, but $p(z|x,y)$ is also intractable.

A (conditional on $y$) variational
autoencoder~\cite{Kingma:2013,rezende2014stochastic} is a deep
learning tool for treating such a latent variable model. It introduces
a \emph{recognition} (or \emph{encoder}) model $q(z|x,y)$, which is
an approximation to the posterior $p(z|x,y)$.  As with the first two
networks, the recognition network should have tractable density and be
easy to sample, e.g., a multivariate Gaussian. One can then take the
expectation (over $z$) of the logarithm of the posterior,
\begin{align}
  \log p(x|y) ={}& \mathbb{E}_{q(z|x,y)} \log p(x|y) \nonumber\\
  ={}& \mathbb{E}_{q(z|x,y)} \log \frac{p(x,z|y)}{q(z|x,y)} \nonumber\\
                 &+ D_{\text{KL}}(q(z|x,y) \| p(z|x,y)) \nonumber\\
  \equiv{}& \mathcal{L} + D_{\text{KL}}(q(z|x,y) \| p(z|x,y)),
\end{align}
where the last term is the KL divergence,
\begin{equation}
  D_{\text{KL}}(q(z|x,y) \| p(z|x,y)) \equiv \mathbb{E}_{q(z|x,y)} \log \frac{q(z|x,y)}{p(z|x,y)}.
\end{equation}
Since this is nonnegative, $\mathcal{L}$ is
known as the \emph{variational lower bound} on $\log p(x|y)$. If
$q(z|x,y)$ is identical to $p(z|x,y)$, then
$\mathcal L = \log p(x|y)$.

The variational autoencoder maximizes the expectation of $\mathcal L$
over the true distribution. The associated loss function can be
written
\begin{align}\label{eq:CVAEloss}
  L_{\text{CVAE}} &= \mathbb{E}_{p_{\text{true}}(x)} \mathbb{E}_{p_{\text{true}}(y|x)} [ - \mathcal{L}] \nonumber\\
              &= \mathbb{E}_{p_{\text{true}}(x)} \mathbb{E}_{p_{\text{true}}(y|x)} \left[ - \mathbb{E}_{q(z|x,y)} \log p(x|z,y) \right.\nonumber\\
              &\qquad\qquad\qquad \left. + D_{\text{KL}}(q(z|x,y) \| p(z|y) ) \right].
\end{align}
The three networks, $p(z|y)$, $p(x|z,y)$, and $q(z|x,y)$, are trained
simultaneously. To evaluate the loss, the outer two expectation values
are treated the same as in the previous subsection. The inner
expectation value is evaluated using a Monte Carlo approximation,
typically with a single sample from $q(z|x,y)$. For Gaussian
$q(z|x,y)$ and $p(z|y)$ the KL divergence term may be calculated
analytically; otherwise, a single Monte Carlo sample suffices.

For training, it is necessary to take the gradient of the loss with
respect to network parameters. The stochasticity of the Monte Carlo
integral estimates must, therefore, be carefully treated. This can be
done by using the \emph{reparametrization
  trick}~\cite{Kingma:2013,rezende2014stochastic}, namely by treating
the random variable as an additional input to the network drawn from a
fixed distribution. For example, if $q(z|x,y)$ is multivariate
Gaussian with mean $\mu(x,y)$ and Cholesky matrix $A(x,y)$, then
\begin{equation}
  z = \mu(x,y) + A(x,y)\epsilon, \qquad \text{with } \epsilon \sim \mathcal{N}(0,1)^l,
\end{equation}
is a sample from $q(z|x,y)$. With this trick, one can now take the
gradient of $z$ with respect to network parameters.

This setup is called a variational \emph{autoencoder} because the
first term in the loss function has the form of an autoencoder. The
recognition network $q(z|x,y)$ is known as the encoder, and $p(x|z,y)$
as the decoder. This first term (the reconstruction loss) is minimized
if $x$ is recovered after being encoded into $z$ and then decoded. The
other term in the loss function (the KL loss) pushes the encoder to
match the prior $p(z|y)$ and acts as a regulator. When the variational
autoencoder works as an autoencoder, the latent variables $z$ can give
a useful low-dimensional representation of $x$.

The recent work~\cite{Gabbard:2019rde} used a CVAE with diagonal
Gaussian networks to model gravitational-wave posterior distributions,
achieving excellent results over the four parameters
$(m_1, m_2, t_c, d_L)$. In the following two subsections we describe
the use of masked autoregressive flows to build even more general
distributions.

\subsection{Masked autoregressive flows}\label{sec:maf}

In this subsection we review the concept of a masked autoregressive
flow, a type of normalizing flow that we use in our work to map simple
distributions into more complex ones. We refer the reader
to~\cite{papamakarios2017masked} for a much more in-depth discussion.

Consider a probability density $p(x)$. Without any loss of generality,
this may be written using the chain rule as
\begin{equation}
  p(x) = \prod_{i=1}^n p(x_i | x_{1:i-1}).
\end{equation}
An autoregressive model restricts each conditional distribution in the
product to have a particular parametrized form. We will take this to
be univariate normal~\cite{papamakarios2017masked},
\begin{equation}
  p(x_i | x_{1:i-1}) = \mathcal{N}(\mu_i(x_{1:i-1}), \exp (2 \alpha_i(x_{1:i-1}) )),
\end{equation}
for $i = 1, \ldots, n$.

In~\cite{kingma2016improved} it was observed that an autoregressive
model defines a normalizing flow. In other words, suppose
$u \sim \mathcal{N}(0,1)^n$. Then
\begin{equation}\label{eq:f}
  x_i = \mu_i(x_{1:i-1}) + u_i \exp \alpha_i(x_{1:i-1})
\end{equation}
gives a sample from $p(x)$. This mapping $f: u \mapsto x$ is defined
recursively in~\eqref{eq:f}, but the inverse mapping,
\begin{equation}\label{eq:finv}
  u_i = \left[x_i - \mu_i(x_{1:i-1})\right] \exp \left( - \alpha_i(x_{1:i-1}) \right),
\end{equation}
is nonrecursive. Because $f$ is autoregressive, the Jacobian determinant is very simple,
\begin{equation}
  \left| \det \frac{\partial(f^{-1}_1,\ldots,f^{-1}_n)}{\partial(x_1,\ldots,x_n)} \right|
= \exp\left( -\sum_{i=1}^n \alpha_i(x_{1:i-1}) \right).
\end{equation}
Hence, $f$ defines a normalizing flow. Starting from a simple base
distribution, the change of variables
rule~(\ref{eq:changeofvariables}) can be used to evaluate the density
$p(x)$.

An autoregressive flow may be modeled by a neural network with
masking~\cite{germain2015made}. That is, one starts with a fully
connected network with $n$ inputs, $n$ outputs, and several hidden
layers to define $f$, but then carefully sets certain connections to
zero such that the autoregressive property holds. This defines a MADE
network~\cite{germain2015made}. Because the inverse direction is
nonrecursive, this requires just a single pass through the
network~\cite{kingma2016improved}. To map in the forward direction
requires $n$ passes. During training it is, therefore, preferable to
only require inverse passes. For gravitational waves, the parameter
space has reasonably low dimensionality, so even the forward pass is
not too expensive.

With a single MADE network, the first component $x_1$ is independent
of all other components, and follows a fixed Gaussian distribution. To
achieve sufficient generality, it is necessary to stack several MADE
networks in sequence, permuting the order of the components between
each pair~\cite{kingma2016improved}. This is called a masked
autoregressive flow (MAF)~\cite{papamakarios2017masked}. For stability
during training it is also useful to insert batch normalization
layers~\cite{ioffe2015batch} between the MADE layers, and between the
base distribution and the first MADE layer; in the context of a flow,
these also contribute to the
Jacobian~\cite{papamakarios2017masked}. MAFs and related networks have
been used to model very complex distributions over high-dimensional
spaces, including audio~\cite{Oord:2016aiw} and
images~\cite{van2016conditional}.

In the context of gravitational-wave parameter estimation, the sample
space is relatively low-dimensional. To model the posterior
distribution, however, each MADE flow unit should be made conditional
on the (high-dimensional) strain data $y$, while maintaining the
autoregressive property over $x$. We can then take a standard normal
base distribution, and flow it through all the MADE and batch
normalization layers, to obtain the complex posterior. The loss
function is the same as in section~\ref{sec:nn-gw}, but with the
change of variables rule used to evaluate the density, i.e.,
\begin{align}\label{eq:mafloss}
  L ={}& \mathbb{E}_{p_{\text{true}}(x)} \mathbb{E}_{p_{\text{true}}(y|x)} \left[ - \log \mathcal{N}(0, 1)^n(f^{-1}(x)) \right. \nonumber\\
    & \qquad \left.  - \log \left| \det \frac{\partial(f^{-1}_1,\ldots,f^{-1}_n)}{\partial(x_1,\ldots,x_n)} \right|
  \right],
\end{align}
where now $f$ denotes the entire sequence of flows in the network.
Notice that this involves only the inverse flow $f^{-1}$, which is
fast to evaluate. This network is fast to train, but somewhat slower
to sample.

\subsection{Combined models}\label{sec:combined}

More powerful models can be obtained by combining autoregressive flows
with the variational autoencoder. Each of the three networks
comprising the CVAE---the encoder $q(z|x,y)$, the decoder $p(x|z,y)$,
and the prior $p(z|y)$---can be made more flexible by applying
autoregressive flows to their sample spaces. We discuss each of these
possibilities in turn. In our experiments, we found that the best
performance was achieved when combining all three. In all cases, the
CVAE loss function~\eqref{eq:CVAEloss} is optimized, with the change
of variables rule used to evaluate the component densities.

\subsubsection{Encoder with inverse autoregressive flow}\label{sec:iaf}

Normalizing and autoregressive flows were first proposed as a way to
increase the flexibility of the encoder
network~\cite{rezende2015variational,kingma2016improved}. This is
important because the CVAE loss function~\eqref{eq:CVAEloss} differs
from the desired cross-entropy loss~\eqref{eq:Hloss} by the
expectation of the KL divergence between $q(z|x,y)$ and the
intractable $p(z|x,y)$. If this can be made smaller, then the two
losses converge, hence $q(z|x,y)$ should be as flexible as possible.

A flexible encoder is also desired to avoid a situation called
``posterior collapse.'' The reconstruction and KL loss terms are in
competition during training, and if the KL loss term collapses to
zero, the network can fail to autoencode. In this situation, the
encoder matches the prior, so it ignores its $x$ input; the latent
variables $z$ contain no information about $x$ beyond what is
contained in $y$. This can happen either because the loss gets stuck
in an undesired local minimum during training, or the configuration
with vanishing KL loss is actually the global
minimum~\cite{chen2016variational}. The former can be alleviated by
careful training strategies such as annealing the KL loss
term~\cite{bowman2015generating}. The latter can occur because the use
of latent variables incurs a cost related to
$D_{\text{KL}}(q(z|x,y)\|p(z|x,y))$~\cite{chen2016variational}. If the
decoder is sufficiently powerful such that it can model $p(x|y)$ on
its own, then $p(z|x,y) \to p(z|y)$ and $L_{\text{CVAE}} \to L$ is the
global minimum. The network simply decides it is not worthwhile to use
the latent variables.

Thus, a flexible encoder is important for performance and to make full
use of latent variables. Since evaluation of $L_{\text{CVAE}}$
requires sampling from $q(z|x,y)$, to map the Gaussian into a more
complex distribution an \emph{inverse autoregressive flow} (IAF)
should be used~\cite{kingma2016improved}. Sampling is fast since the
inverse flow \eqref{eq:finv} does not involve recursion. The density
evaluation needed for the KL loss term only needs to take place for
$z$ sampled from $q(z|x,y)$, so caching may be used.

Each MADE layer comprising the encoder IAF may be conditioned on $x$,
$y$, and an optional additional output of the Gaussian encoder,
$h$. (The initial work~\cite{kingma2016improved} used only $h$.) In
our experiments we found it most effective to condition only on $h$
and $x$. It is possible that the high dimensionality of $y$ meant that
it overpowered $x$.

\subsubsection{Decoder with masked autoregressive flow}

Since the reconstruction loss requires density evaluation of the
decoder distribution, one should apply a forward MAF after the
Gaussian distribution to increase flexibility of $p(x|z,y)$. The MADE
layers may be conditioned on $y$ (as in the basic MAF of
section~\ref{sec:maf}) and the latent variable $z$. This powerful
decoder increases the risk of posterior collapse, so it is useful to
use also the IAF encoder.

\subsubsection{Prior with masked autoregressive flow}

As described in~\cite{chen2016variational}, an autoregressive flow at
the end of the prior network $p(z|y)$ effectively adds flexibility to
the encoder network and the decoder network. For fast training, one
should again use a forward MAF, and its MADE layers should be
conditioned on $y$.

Although a prior MAF and an encoder IAF are very closely related, they
differ in that the IAF can be conditioned on $x$. We found that this
had a significant impact on performance, and therefore included both
autoregressive flows in our models.

\section{Nonspinning binaries}\label{sec:case1}

In this section we describe our experiments in using deep learning
models to describe nonspinning coalescing black hole
binaries. Binaries are described by five parameters: the masses $m_1$
and $m_2$, the luminosity distance $d_L$, the time of coalescence
$t_c$, and the phase of coalescence $\phi_0$. These parameters and
their ranges are chosen to facilitate comparison
with~\cite{Gabbard:2019rde}.

\subsection{Training data}\label{sec:data}

Training data for our models consist of pairs of parameters $x$ and
strain time series $y$. Parameters are sampled from a prior
distribution $p(x)$, which is uniform over each parameter except for the
volumetric prior over $d_L$. Parameter ranges are taken
from~\cite{Gabbard:2019rde},
\begin{subequations}
  \begin{IEEEeqnarray}{rCl}
  35~\mathrm{M_\odot} \le{}& m_1, m_2 &{}\le 80~\mathrm{M_\odot},\\
  1000~\mathrm{Mpc} \le{}& d_L &{}\le 3000~\mathrm{Mpc}, \\
  0.65~\mathrm{s} \le{}& t_c &{}\le 0.85~\mathrm{s}, \\
  0 \le{}& \phi_0 &{}\le 2\pi.
  \end{IEEEeqnarray}
\end{subequations}
We also take $m_1 \ge m_2$.

Strain realizations are in the time domain and 1~s long
($0\le t \le 1~\mathrm{s}$) with a sampling rate of
$1024~\mathrm{Hz}$. (We found that the sampling rate of
$256~\mathrm{Hz}$ of~\cite{Gabbard:2019rde} was not sufficient to
eliminate Gibbs ringing.) Strain data consist of a waveform $h(x)$,
deterministic from parameters $x$, and random noise $n$, sampled from
the Advanced LIGO~\cite{TheLIGOScientific:2014jea} Zero Detuned High
Power PSD~\cite{zdhp}. We took $h(x)$ to be the ``$+$'' polarization
of \verb+IMRPhenomPv2+ waveforms~\cite{Khan:2015jqa}, which,
following~\cite{Gabbard:2019rde}, we whitened in the frequency domain
using the PSD before taking the inverse Fourier transform to the time
domain. Finally, we rescaled our waveforms by dividing by
$\sqrt{2\Delta t}$ so that in the end, time domain noise is described
by a standard normal distribution in each time sample, i.e.,
$n\sim \mathcal{N}(0,1)^m$.

The whitening and rescaling procedure serves two purposes: it means
that we can easily draw noise realizations at train time by sampling
from a standard normal distribution, and it ensures that the input
data to the neural network has approximately zero mean and unit
variance. The latter condition (``standardization'' of data) has been
shown to improve training performance. Similarly, we rescale $x$ to
have zero mean and unit variance in each component across the training
set.

\begin{figure}
  \includegraphics[width=\columnwidth]{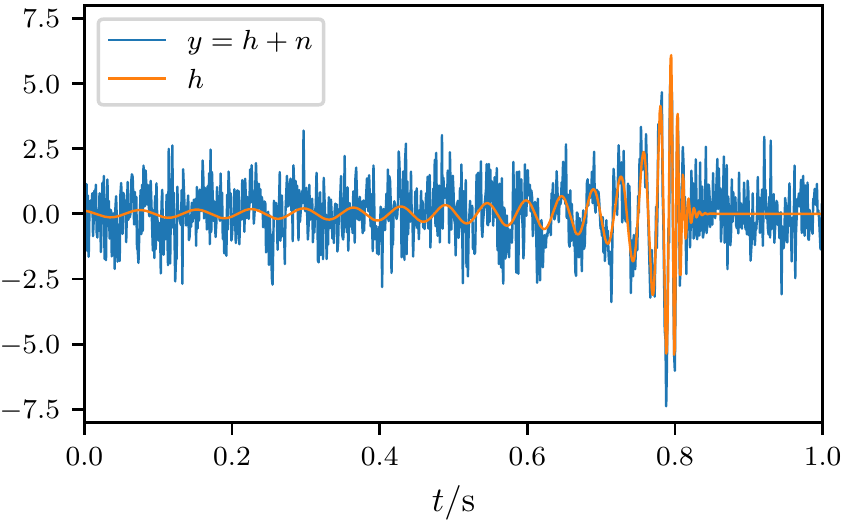}
  \caption{\label{fig:sample_waveform}Sample waveform $h$ and strain
    realization $y$. The SNR for this injection is
    24.9.}
\end{figure}

\begin{figure}
  \includegraphics[width=\columnwidth]{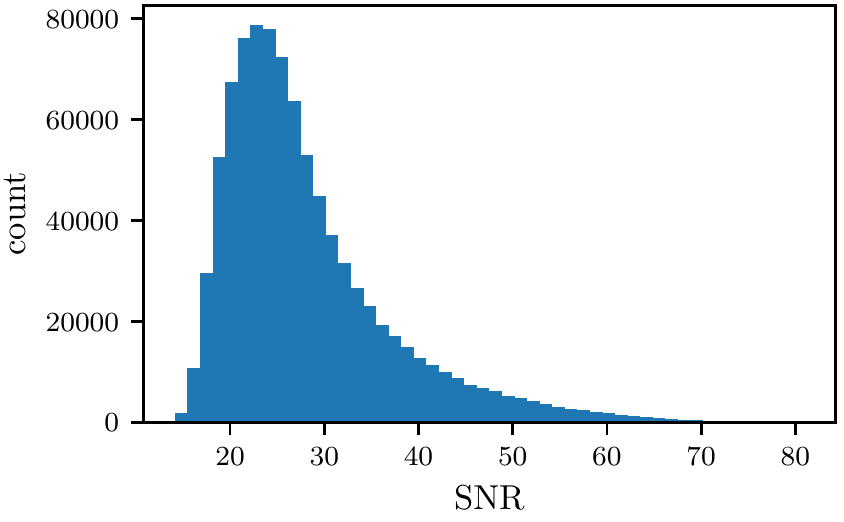}
  \caption{Histogram of SNR values for our training set.\label{fig:snr5D}}
\end{figure}

Our dataset consists of $10^6$ $(x,h)$ pairs, which we split into 90\%
training data and 10\% validation data. Noise realizations $n$ are
sampled at train time to give strain data $y=h+n$; a sample time
series is given in figure~\ref{fig:sample_waveform}. The median
signal-to-noise ratio (SNR) of our training set is 25.8, and the
complete SNR distribution is given in given in
figure~\ref{fig:snr5D}. Our dataset is the same size as that
of~\cite{Gabbard:2019rde} and a factor of $10^3$ smaller than the
effective dataset of~\cite{Chua:2019wwt}. Larger datasets are
generally preferred to prevent overfitting, but they are also more
costly to prepare and store. We would like to build a system that can
generalize well with as small a dataset as possible in order to
minimize these costs when moving to more complicated and longer
waveforms in the future. By keeping track of performance on the
validation set when training our models, we found that our training
set was sufficiently large to avoid overfitting. We also experimented
with a $10^5$-element dataset; this had slightly reduced, but still
acceptable, performance.

\subsection{Experiments}\label{sec:exp}

\begin{table}
  \caption{\label{modelcompcost}Network depth, number of trainable
    parameters, and training time. Numbers of trainable parameters are
    approximated as the numbers of weights; for autoregressive flows,
    we included an estimate of the number of \emph{unmasked} weights.}
  \begin{ruledtabular}
    \begin{tabular}{l c c c}
      Model & Layers & Parameters ($\times10^6$) & Train time (h)\\
      \hline
      CVAE & 12 & 31.6 & 5.6 \\
      MAF & 20 & 4.0 & 6.5 \\
      CVAE+ & 72 & 18.8 & 14.8 \\
    \end{tabular}
  \end{ruledtabular}
\end{table}
\begin{table}
  \caption{\label{modellosses}Final values of training and validation
    set loss functions after 250 epochs.}
  \begin{ruledtabular}
    \begin{tabular}{lcccc}
      Model & Train loss & Test loss & KL train loss & KL test loss\\
      \hline
      CVAE & $-4.52$ & $-4.31$ & $4.33$ & $4.49$ \\
      MAF & $-4.43$ & $-4.42$ & -- &  -- \\
      CVAE+ & $-7.00$ & $-6.95$ & $6.76$ & $6.77$ \\
    \end{tabular}
  \end{ruledtabular}
\end{table}

We modeled our gravitational-wave posteriors with three types of
neural network described in section~\ref{sec:prelim}: a CVAE (similar
to~\cite{Gabbard:2019rde}), a MAF, and a CVAE with autoregressive
flows appended to the three subnetworks (denoted CVAE+). We selected
hyperparameters based on choices in the
literature~\cite{papamakarios2017masked,Gabbard:2019rde} and on sweeps
through hyperparameter space. Approximate network sizes and training
times for each architecture are listed in table~\ref{modelcompcost},
and final loss functions after training in table~\ref{modellosses}.

All of our networks use ReLU nonlinearities. They were trained for 250
epochs with a batch size of 512, using the Adam
optimizer~\cite{Kingma:2014vow}. The learning rate was reduced by a
factor of two every 80 epochs. Sampling performance results for the
three network architectures are collected in figures \ref{corner} and
\ref{pp}.

\subsubsection{CVAE}\label{sec:cvae}

Our CVAE network is designed to be similar to that
of~\cite{Gabbard:2019rde}. The encoder, decoder, and prior
distributions are all taken to be diagonal Gaussians, each
parametrized by fully connected neural networks with three hidden
layers of dimension 2048. The latent space is of dimension $l=8$. We
used the same initial training rate as~\cite{Gabbard:2019rde}, of
0.0001.

\begin{figure*}
  \subfloat[][\label{cornercvae}CVAE]{\includegraphics[width=\columnwidth]{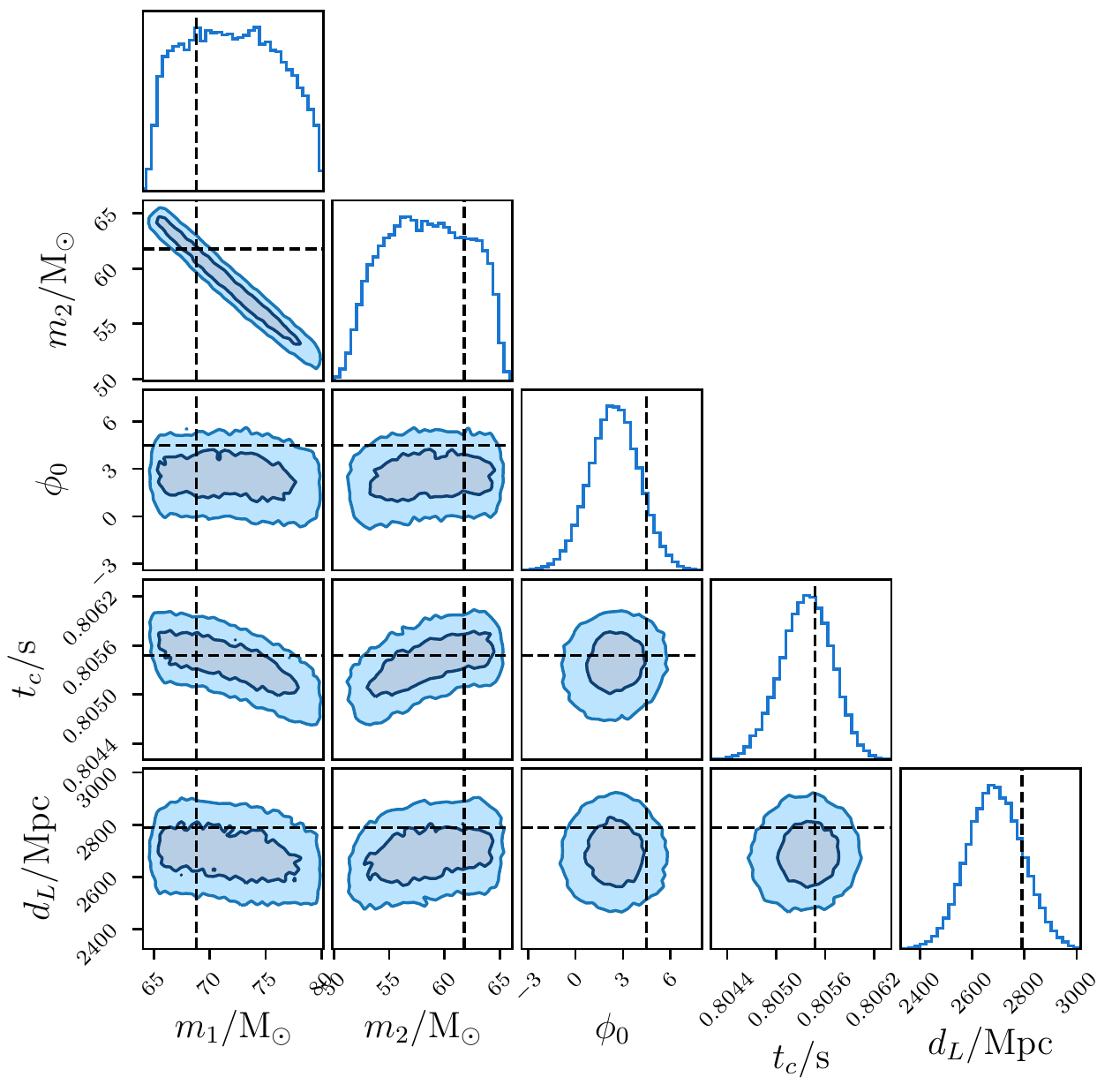}}
  \hfill
  \subfloat[][\label{cornermaf}MAF]{\includegraphics[width=\columnwidth]{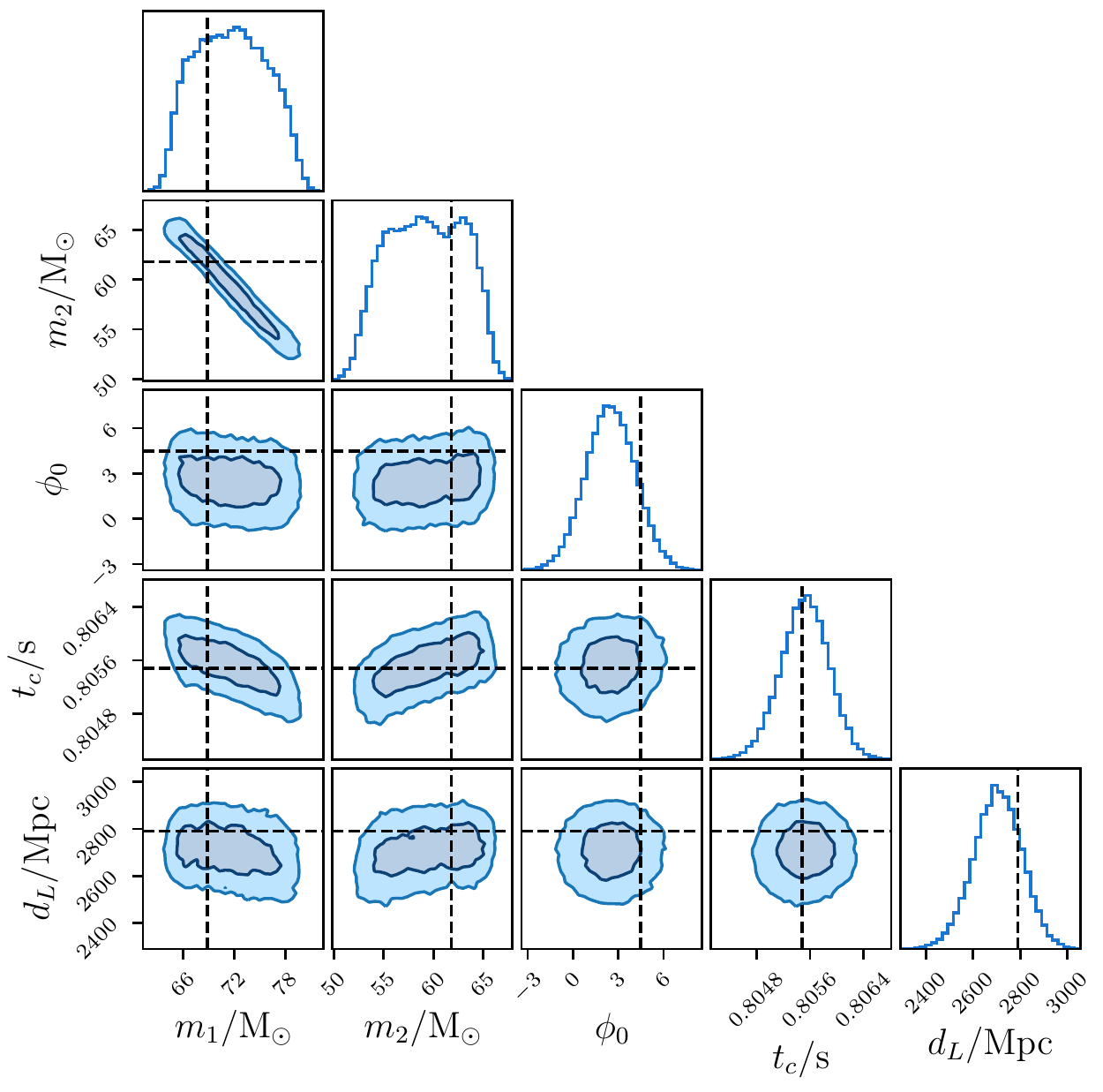}}
  
  \subfloat[][\label{fig:cornercvaeall}CVAE+]{\includegraphics[width=1.5\columnwidth]{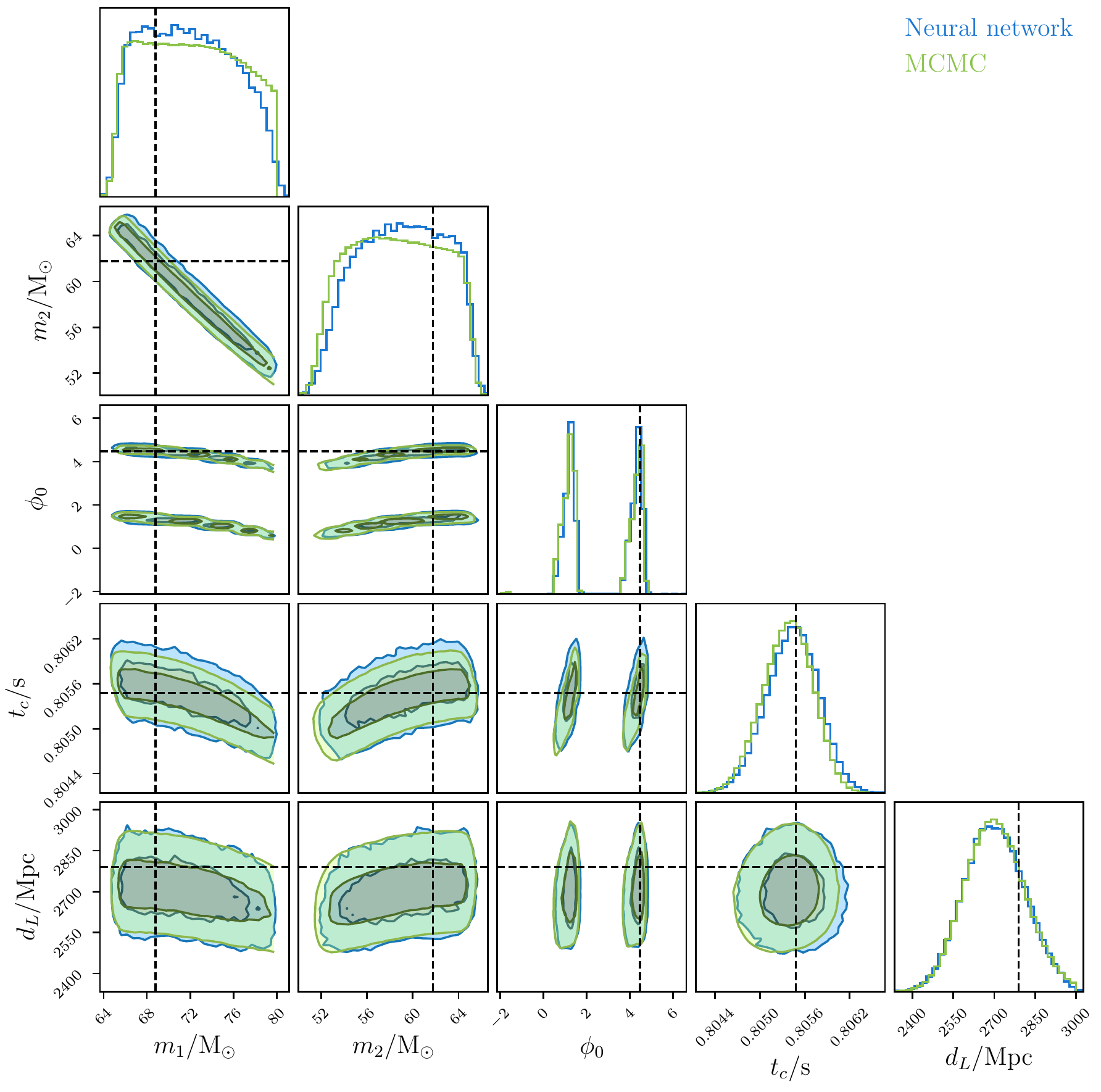}}
  \caption{\label{corner}One- and two-dimensional marginalized
    posterior distributions for the strain realization shown in
    figure~\ref{fig:sample_waveform}, comparing output from CVAE, MAF,
    and CVAE+ neural networks. Each figure is constructed from
    $5\times10^4$ samples, and contours represent 50\% and 90\%
    credible regions. For the CVAE+ network, MCMC results are
    overlayed for comparison. The CVAE+ network is the only one
    capable of capturing the multimodality in the $\phi_0$ posterior.}
\end{figure*}

\begin{figure*}
  \subfloat[][\label{cvaepp}CVAE]{\includegraphics[width=\columnwidth]{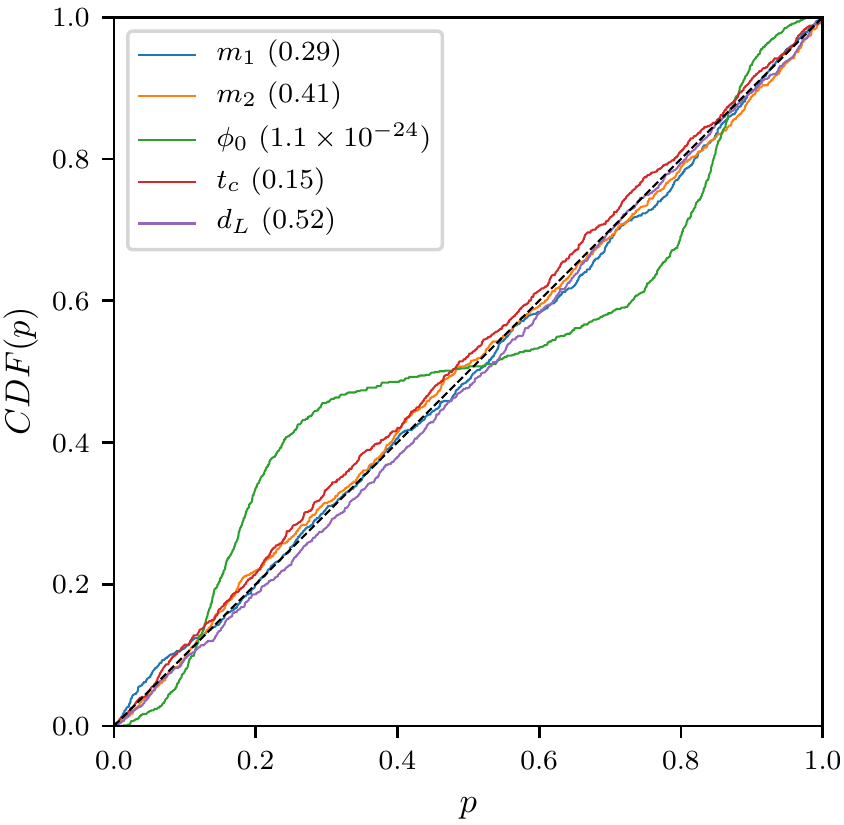}}
  \hfill
  \subfloat[][\label{mafpp}MAF]{\includegraphics[width=\columnwidth]{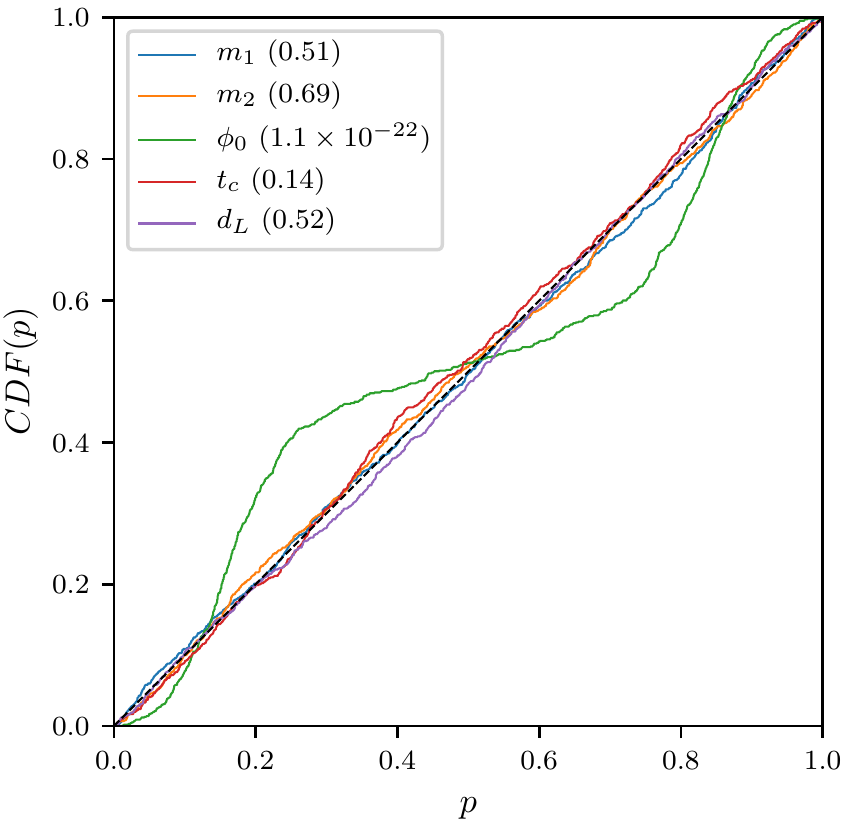}}
  
  \subfloat[][\label{cvae+pp}CVAE+]{\includegraphics[width=\columnwidth]{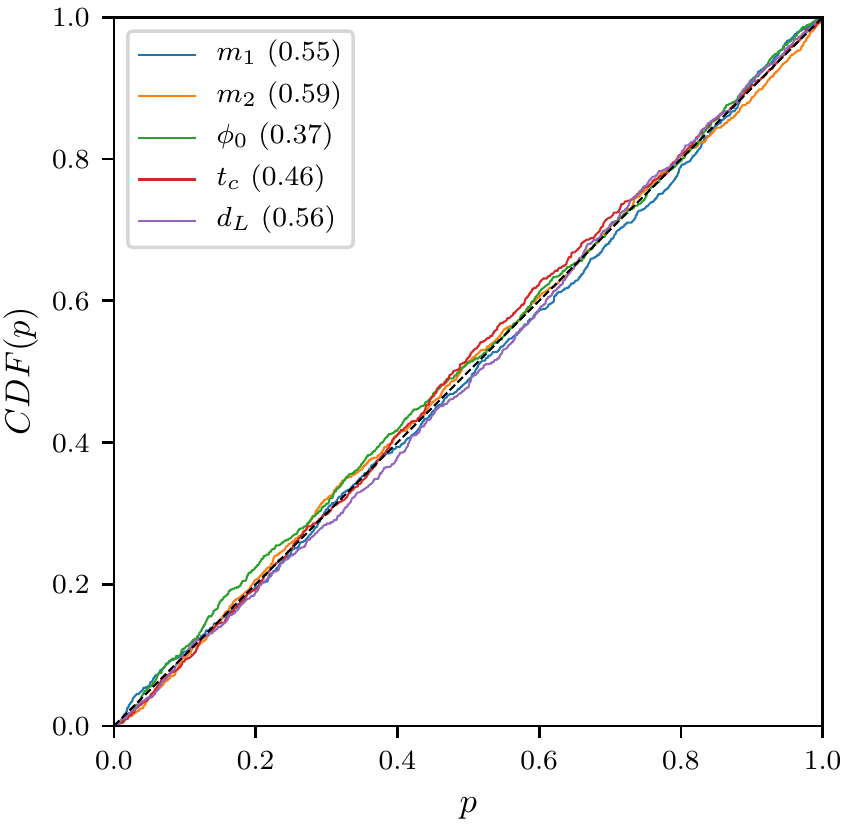}}
  \hfill
  \caption{\label{pp}p--p plots for the three neural network
    models. For each marginalized one-dimensional posterior
    distribution, the curve represents the cumulative distribution
    function of the percentile scores of the true values of the
    parameter. Each plot is constructed from the same $10^3$ parameter
    choices and strain realizations. Deviations from the diagonal
    represent a failure of the model to learn the true posterior for
    that parameter. KS test $p$-values are provided in the legends.}
\end{figure*}

To train the CVAE, we optimize the variational lower bound loss
$L_{\text{CVAE}}$ of~\eqref{eq:CVAEloss}, with a single Monte Carlo
sample to estimate the expectation value over $q(z|x,y)$. Since
$L_{\text{CVAE}}$ is only an upper bound on the cross-entropy loss
$L$, the value of the loss function alone is not entirely indicative
of performance. Indeed, when posterior collapse occurs (see
section~\ref{sec:iaf}), the gap between $L_{\text{CVAE}}$ and $L$ can
vanish; in this case, the value of $L$ is larger than that of a
network with the same $L_{\text{CVAE}}$ and no posterior collapse. For
this reason, it is important to also use other metrics to evaluate
performance. For CVAE models we always quote the KL loss as an
indication that the latent space is being used.

To encourage use of the latent space, we found it to be beneficial to
use KL annealing during training, i.e., we multiplied the KL loss part
of $L_{\text{CVAE}}$ by a factor between zero and one during the early
stages. This reduces the importance of the KL loss term compared to
the reconstruction loss. For all CVAE-based moels, we adopted a cyclic
KL annealing strategy~\cite{fu2019cyclical}: for the first 6 epochs we
used annealing factors of $(10^{-5},1/3,2/3,1,1,1)$, and we repeated
this cycle 3 more times; see figure~\ref{fig:tvcvae}. We also ignored
the KL loss term whenever it was less than 0.2.

In figure~\ref{cornercvae}, we show a posterior distribution
corresponding to the strain data $y$ of
figure~\ref{fig:sample_waveform}. This is constructed from $N=5\times10^4$
posterior samples, obtained using formula~\eqref{eq:posterior-latent}
with the prior and decoder networks as follows: (1) pass $y$ through
the prior network to obtain the distribution $p(z|y)$, (2) draw $N$
latent-space samples $z^{(i)}$ from the prior, (3) pass these through
the decoder network to obtain $N$ distributions $p(x|z^{(i)},y)$,
and finally (4) draw one sample $x^{(i)}$ from each of these
distributions.

By inspection of the posterior, it is clear that the latent space is
being used in the model, since the distribution is not a diagonal
Gaussian. The distributions for most of the parameters look
reasonable, and they cover the true values of the parameters.  The
phase of coalescence, $\phi_0$, is, however, not being captured at
all. Indeed, $\phi_0$ should be precisely $\pi$-periodic because our
training set waveforms only contain the $(l,m) = (2,2)$ mode of the
signal. Moreover, since we are taking a single polarization, $\phi_0$
should be resolvable.

We can evaluate the performance of the network with a p--p
plot~\cite{Veitch:2014wba}. To do this, we compute posterior
distributions, each comprised of $10^4$ samples, for $10^3$ different
strain realizations (i.e., $y=h(x) + n$ for $x$ drawn from the prior
over parameters and $n$ a noise realization). For each
one-dimensional posterior, we then compute the percentile value of the
true parameter. For each parameter, the p--p curve is then the
cumulative density function of the percentile values. This is shown in
figure~\ref{cvaepp} for the CVAE network. If the CDF is diagonal, then
the percentile values are distributed according to a uniform
distribution, as one would expect for the true one-dimensional
posteriors. We can see that the CVAE appears to capture all of the
one-dimensional distributions except for $\phi_0$.

To confirm that the percentile scores are well-distributed, we also
performed a Kolmogorov-Smirnov test. We calculated the KS statistic to
compare the distribution of percentile scores to a uniform
distribution. We found that $\phi_0$ had miniscule $p$-value, as
expected, the $p$-value of $t_c$ was 0.15, and all other $p$-values
were greater than 0.29. This indicates that all parameters are
well-recovered, except for $\phi_0$.

\subsubsection{MAF}\label{sec:maf}

Next, we modeled the gravitational-wave posterior $p(x|y)$ directly
using a masked autoregressive flow. As described in
section~\ref{sec:maf}, the MAF network describes the posterior in
terms of an invertible mapping $f$ from a five-dimensional standard
normal distribution $\mathcal{N}(0,1)^5(u)$ into the
gravitational-wave posterior $p(x|y)$. The flow $f(u,y)$ is
autoregressive over $u$ and has arbitrary dependence on strain data
$y$. The MAF network does not involve latent variables, and optimizes
the cross-entropy with the data distribution. Thus the loss function
$L$ [given in~\eqref{eq:mafloss}] can be used to directly compare
performance for different models.

Our best performing MAF network consists of five MADE units, each with
three hidden layers of dimension 512, and conditioned on $y$. We also
inserted batch normalization layers between each pair of MADE units,
and between the first MADE unit and the base space. We found this to
be important for training stability. We were able to train the MAF
successfully at a higher initial learning rate than the CVAE, of
0.0004. Following~\cite{papamakarios2017masked}, at the end of each
training epoch, before computing the validation loss, we set the
stored mean and variance vectors of the batch normalization layers to
the mean and variance over the entire training set. (We did this also
for the CVAE+ network below.)

In figure~\ref{cornermaf}, we show a corner plot for the same strain
data as before. Samples are obtained by first sampling from the
standard normal base space, $u \sim \mathcal N(0,1)^5$ and then
applying the the flow $f(u,y)$. All quantities appear to be
well-modeled except, again, for $\phi_0$.\footnote{In some experiments
  with deeper MAF networks, we were in fact able to properly resolve
  the $\phi_0$ posterior. This was, however, difficult to reproduce
  and somewhat unstable because MAF networks use element-wise affine
  transformations~\eqref{eq:f}, and these may not be flexible enough
  to consistently resolve multimodality. We leave further
  investigation to future work.}  This is consistent with the p--p
plot shown in figure~\ref{mafpp}. The KS statistic $p$-values were
0.14 for $t_c$, and they were greater than 0.5 for all other
parameters (except for $\phi_0$). (The p--p plots for all networks are
computed from the same strain realizations, so it is consistent to see
the same KS statistic $p$-values for the different networks.)

Performance of the MAF network is comparable to that of the CVAE, with
(in our case) one eighth the number of trainable parameters. In both
cases, all parameters except for $\phi_0$ are well-modeled by the
networks. In addition, the final loss values given in
table~\ref{modellosses} are very close for both networks, with a
slight edge in validation loss for the MAF. However, since
$L_{\text{CVAE}}$ is an upper bound on $L$, the cross-entropy loss for
the CVAE network may actually be lower than that of the MAF. Indeed,
the fact that the CVAE made use of the latent space suggests that this
gap is nonzero. Comparison of the training and validation loss
functions shows slight overfitting for the CVAE, but none for the MAF.

\subsubsection{CVAE+}\label{sec:cvae+}

Finally, we experimented with combinations of CVAE and MAF
networks. As described in section~\ref{sec:combined}, all three
component distributions of the CVAE can be made more flexible by
applying MAF transformations to the diagonal Gaussian distributions,
thereby increasing the total modeling power of the
network~\cite{kingma2016improved,chen2016variational}. Indeed,
appendix A of~\cite{kingma2016improved} shows that a single linear
autoregressive flow is capable of generalizing a Gaussian distribution
from diagonal to generic covariance.

For the combined models, the initial Gaussian distributions (the base
spaces for the autoregressive flows) are modeled in the same way as in
section~\ref{sec:cvae}, as fully connected three-hidden-layer
networks. However, we reduced the size of hidden layers to 1024
dimensions. We also kept the $l=8$ dimensional latent space. We found
that best performance was achieved by applying autoregressive flows to
all three component distributions as follows:
\begin{description}
\item[$q(z|x,y)$] We added an IAF after the initial Gaussian
  encoder. This was made conditional on $x$ and extra 8-dimensional
  context output $h$ from the initial encoder. We also experimented
  with conditioning on $y$, but found this to reduce performance.
\item[$p(z|y)$] We added a MAF after the initial Gaussian prior
  network. This was made conditional on $y$.
\item[$p(x|z,y)$] We added a MAF after the initial Gaussian decoder
  network. This was conditional on $y$ and $z$.
\end{description}
In all cases the MAF/IAF parts consist of 5 MADE units (three hidden
layers of 512 dimensions each) and batch normalization
layers. Although the CVAE+ network contains far more layers than the
basic CVAE, it has roughly half the number of free parameters since we
reduced the width of the hidden layers in the Gaussian components.

Training was performed by optimizing $L_{\text{CVAE}}$, with the
change of variables rule used to evaluate the component
densities. Sampling was similar to the basic CVAE, but now with the
appropriate flows applied after sampling from the Gaussians.

We found that optimization was best when we trained the Gaussian
distributions at a learning rate of 0.0001, and the autoregressive
networks at a higher rate of 0.0004, combining the rates of the
previous two subsections. As always, the learning rate was reduced by
half every 80 epochs.

\begin{figure}
  \includegraphics[width=\linewidth]{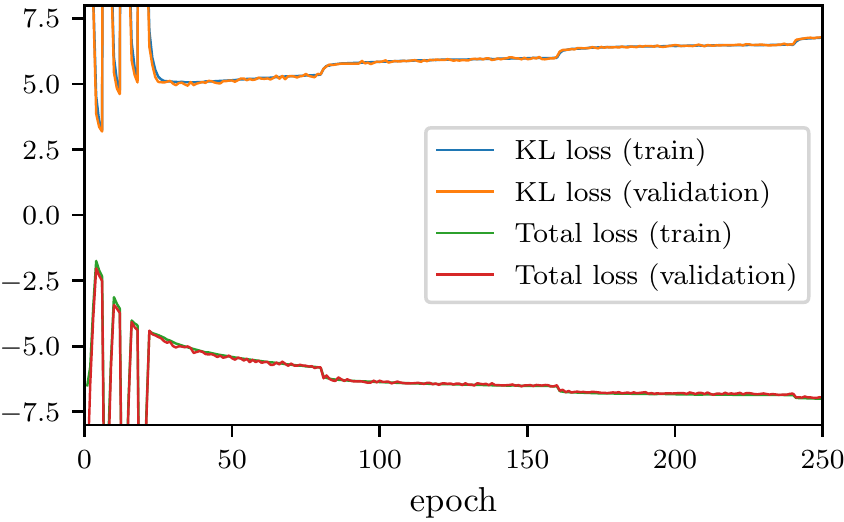}
  \caption{\label{fig:tvcvae}Training and validation loss for each
    epoch for the CVAE+ network. Spikes at early stages arise from the
    cyclic KL annealing. Since training and validation loss coincide,
    there is negligible overfitting.}
\end{figure}

It is especially important for the CVAE+ network to apply some sort
of KL annealing to encourage use of the latent space. An important
difference compared to the basic CVAE of section~\ref{sec:cvae} is
that now the decoder network is sufficiently powerful to model much of
the posterior \emph{on its own}. Indeed, it is just as flexible as the
MAF network of the previous subsection, which produced the posterior
in figure~\ref{cornermaf}, and it is well known that a CVAE will often
ignore latent variables if the decoder is sufficiently
powerful~\cite{bowman2015generating,chen2016variational}. This
strategy combined with the flexible encoder distribution resulted in
higher KL validation loss for the CVAE+ network (6.77) than the basic
CVAE (4.49) by the end of training; see table~\ref{modellosses}. A
plot showing the KL and total loss terms as training proceeds is given
in figure~\ref{fig:tvcvae}.

A sample gravitational-wave posterior distribution for CVAE+ is given
in figure~\ref{fig:cornercvaeall}. In contrast to the simpler models,
this captures the periodicity in $\phi_0$. Sampling was very fast,
requiring $\approx 3.3~\mathrm{s}$ to obtain the $5\times10^4$ samples
that make up this figure. The p--p plot in figure~\ref{cvae+pp} shows
excellent statistical performance, with all KS statistic $p$-values
greater than 0.35.

For validation of our network against standard methods,
figure~\ref{fig:cornercvaeall} also includes samples obtained using
MCMC. We used the emcee ensemble sampler~\cite{Foreman_Mackey_2013}
with the same prior and likelihood as defined by our training set
construction. There is clear agreement between MCMC sampling and
neural network sampling, with slight deviation in the mass
posteriors. We expect that this could be reduced with hyperparameter
improvements or further training. The KL divergence between the two
distributions shown is estimated~\cite{wang2009divergence} at
\begin{align}
  D_{\text{KL}}\left(p_{\text{CVAE}+}(x|y)\| p_{\text{MCMC}}(x|y)\right) &< 0.1,\\
  D_{\text{KL}}\left(p_{\text{MCMC}}(x|y)\| p_{\text{CVAE}+}(x|y)\right) &= 0.1.
\end{align}

\section{Inclined binaries with aligned spins}\label{sec:spins}

To test our method with a more challenging dataset, we augmented the
parameter space to include nonzero aligned spins and inclination
angle. We took uniform priors on $\chi_{1z}$, $\chi_{2z}$, and
$\cos \theta_{JN}$ between $-1$ and $1$, keeping the rest of the prior
distribution unchanged. The neural network was trained to model
posterior distributions over the eight-dimensional parameter space
$x=(m_1, m_2, \phi_0, t_c, d_L, \chi_{1z}, \chi_{2z},
\theta_{JN})$. We held the dataset size fixed at $10^6$ elements,
again split $90\%$ into the training set and $10\%$ into the
validation set. The median SNR for this training set is 17.2, the mean
is 19.1, and the range is $(6.1, 100.9)$.

We tested only the CVAE+ model, since this performed best in previous
experiments. Because gravitational-wave posteriors over the
eight-dimensional parameter space have increased complexity, we
increased the capacity of our network by doubling the latent space
dimension to $l=16$, the number of MADE units to 10 per component
network, and the dimension of the IAF context variable to 16. We also
found performance was best when we froze the Gaussian part of the
prior network, but aside from this all other hyperparameters were
unchanged. This increased the total number of trainable parameters to
$25.2 \times 10^6$.

We trained again for 250 epochs, with final total loss values of
$-0.83$ (train) and $-0.73$ (validation), and final KL loss values of
11.05 (train) and 11.14 (validation), showing very little overfitting.
This also indicates that for the extended parameter space, the network
made heavier use of the latent space. (This observation motivated the
increased size of the latent space.)

\begin{figure*}
  \stackinset{r}{}{t}{}{\includegraphics[width=1.1\textwidth*3/8]{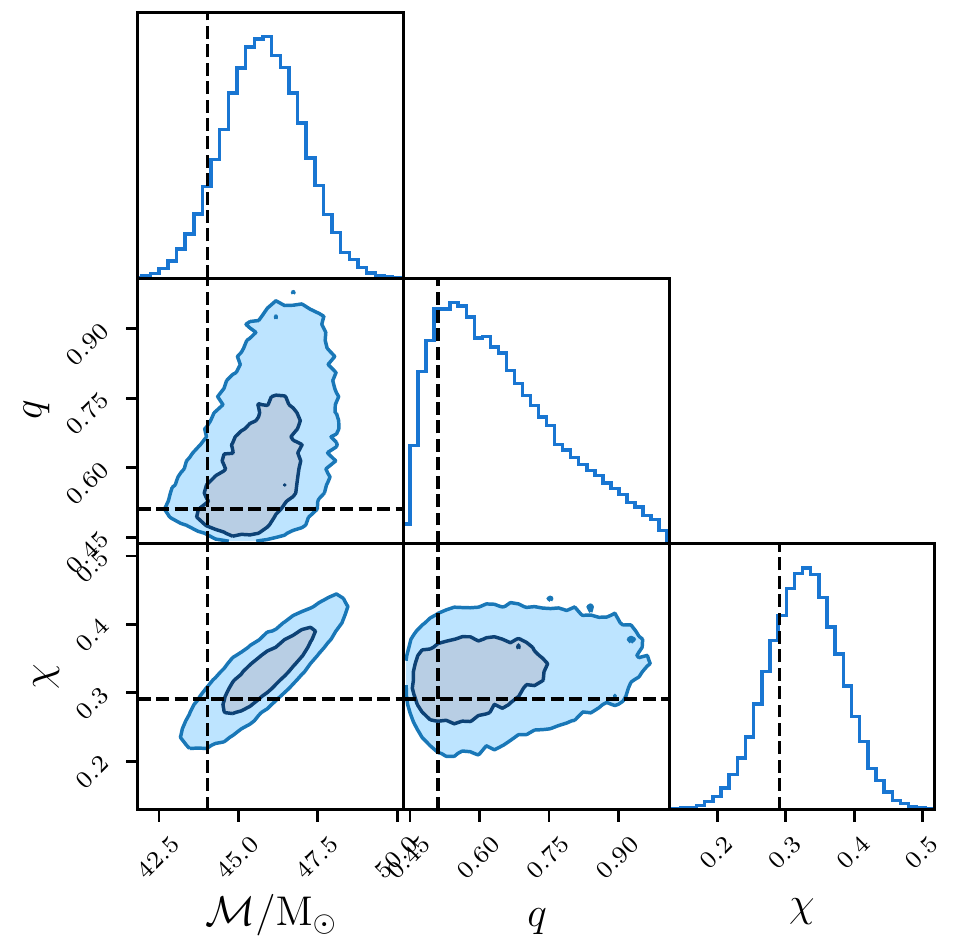}}{
  \includegraphics[width=\textwidth]{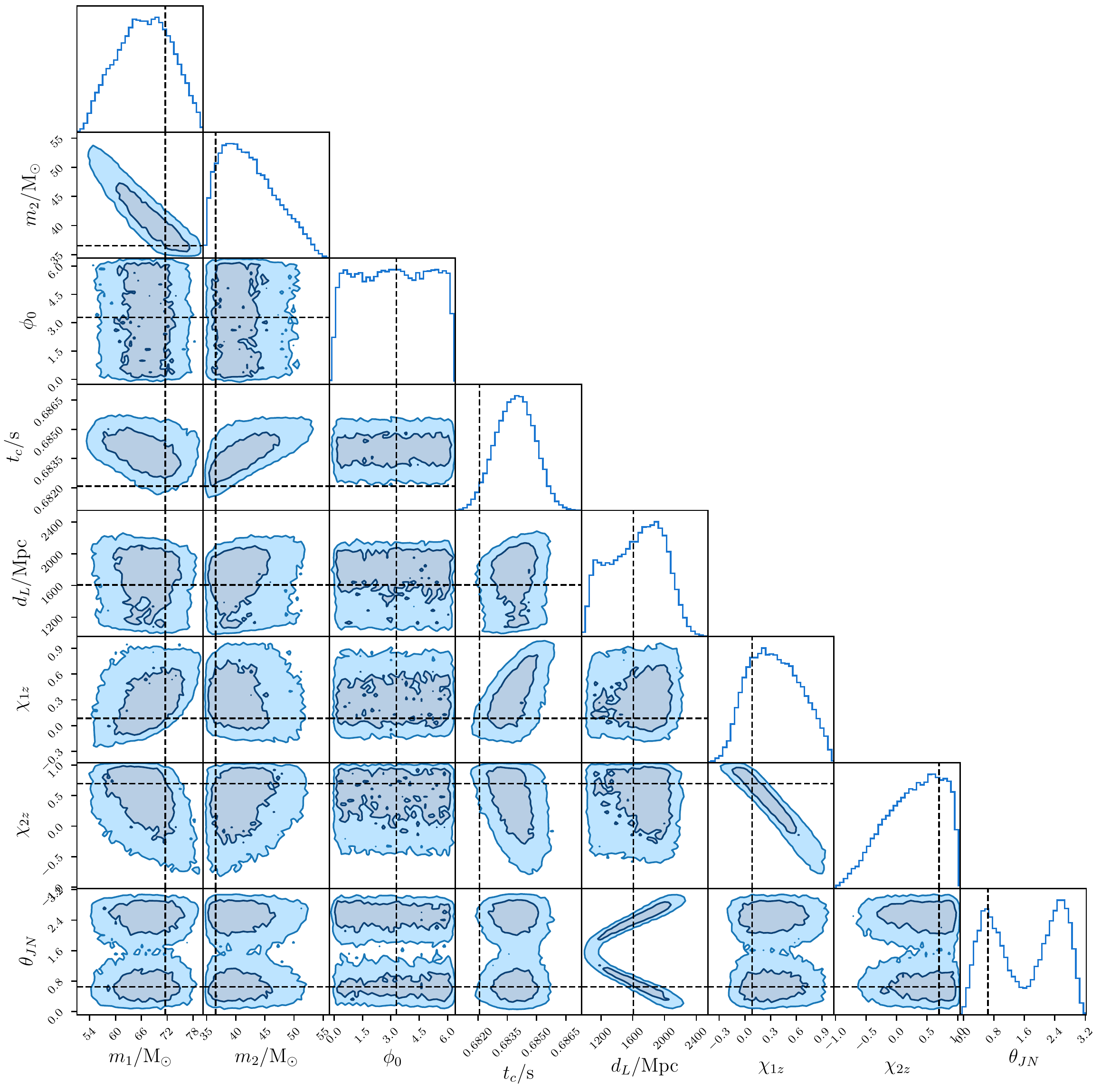}}
\caption{Sample posterior distribution for eight-dimensional parameter
  space, from $5\times10^4$ samples of CVAE+. Inset shows derived
  quantities: chirp mass $\mathcal{M}$, asymmetric mass ratio $q$, and
  effective spin parameter $\chi$. The waveform injection has SNR of
  29.6.\label{fig:extended_posterior}}
\end{figure*}

\begin{figure}
  \includegraphics[width=\linewidth]{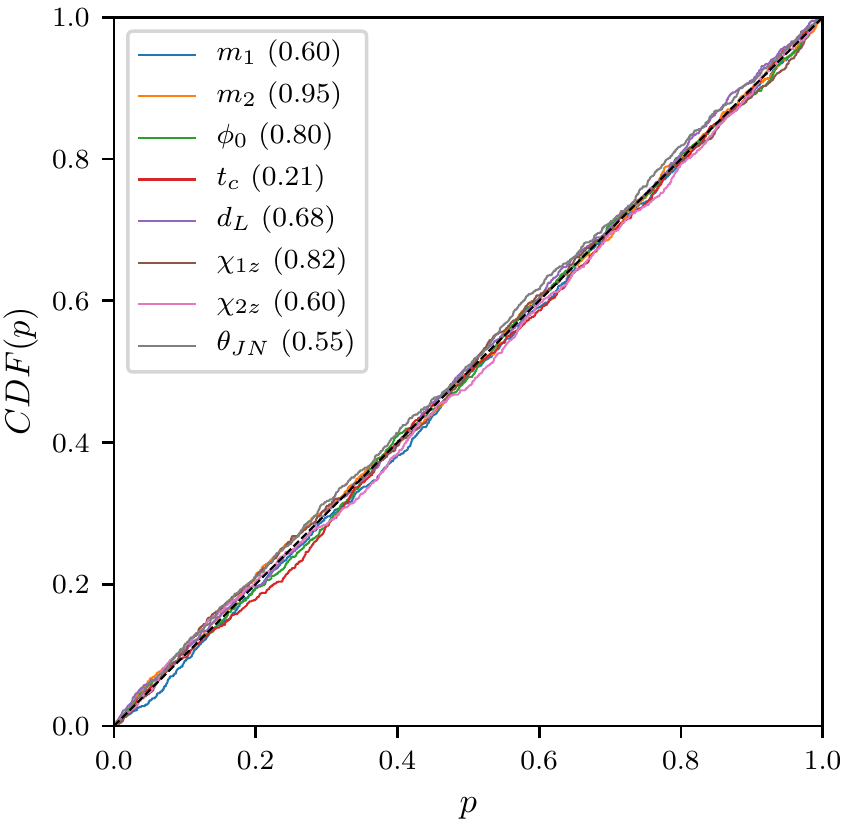}
  \caption{p--p plot for one-dimensional posteriors for
    eight-dimensional parameter space, modeled with CVAE+
    network. Constructed from $10^3$
    injections.\label{fig:extended_pp}}
\end{figure}

In figure~\ref{fig:extended_posterior} we show a representative
posterior distribution produced by the neural network. It continues to
capture the $m_1$--$m_2$ degeneracy, as well as the new
$\chi_{1z}$--$\chi_{2z}$ and $d_L$--$\theta_{JN}$ degeneracies that
arise in the extended parameter space. In contrast to the smaller
parameter space of the previous section, the posterior over $\phi_0$
is simply the prior. The reason for this is that in our waveform
model, there is a three-parameter degeneracy between the two spins
$\chi_{1z}$, $\chi_{2z}$, and the phase $\phi_0$, where small changes
in the spins cause large changes in the phase. We confirmed with MCMC
sampling that the posterior over $\phi_0$ is correct. A p--p plot for
the extended parameter space is presented in figure
\ref{fig:extended_pp}. By inspection, this shows good recovery of all
parameters. Moreover, all parameters have KS $p$-values greater than
0.2.

The inset in figure~\ref{fig:extended_posterior} shows the posterior
distribution over the chirp mass
$\mathcal{M} = (m_1m_2)^{3/5}/(m_1+m_2)^{1/5}$, asymmetric mass ratio
$q = m_2/m_1$, and effective spin parameter
$\chi = (m_1\chi_{1z} + m_2\chi_{2z})/(m_1+m_2)$. These are derived
quantities, with sampling done instead over the parameters listed
above. Although posteriors are simpler over the derived parameters, to
test our method we chose to sample over parameters with more
nontrivial posteriors.

Sampling from the larger CVAE+ model used for the extended parameter
space is slightly slower than the smaller model of the previous
section, now requiring $\approx 1.6~\mathrm{s}$ to obtain $10^4$
posterior samples. This is partially due to the larger 16-dimensional
latent space: the forward pass through a MAF is recursive, so twice as
many passes are required to sample from the variational prior and
decoder. Both MAFs also have twice as many layers.

\section{Conclusions}\label{sec:conclusions}

In this work we introduced the use of masked autoregressive flows for
improving deep learning of gravitational-wave posterior
distributions. These learnable mappings on sample spaces induce
transformations on probability distributions, and we used them to
increase the flexibility of neural network posterior models. The
models that we built can all be rapidly sampled, requiring
$< 2~\mathrm{s}$ to obtain $10^4$ samples.

For nonspinning, quasi-circular binary black holes, and a single
gravitational-wave detector (a five-dimensional parameter space) we
compared models involving a single MAF, a CVAE, and a CVAE with
autoregressive flows applied to the encoder, decoder, and prior
networks (CVAE+). We found that the performance of the single MAF and
CVAE models were comparable, and that best performance was achieved by
the CVAE+ model. The CVAE+ model was able to capture the bimodality in
the phase $\phi_0$, which eluded the other models.

We then considered a larger eight-dimensional parameter space, with
aligned spins and nonzero binary inclination angle. With a
higher-capacity CVAE+ network, we successfully recovered the posterior
distribution over all parameters. This demonstrates that our approach
extends to higher-dimensional parameter spaces. Modest increase in
network capacity may, however, be required.

Although best performance was achieved with the CVAE+ model, an
advantage of models without latent variables (such as the MAF alone)
is that it is possible to directly evaluate the probability density
function. Since the posterior distribution that the MAF models is
normalized, one could then calculate the Bayesian evidence by
separately evaluating the likelihood and prior. (In CVAE+, this would
require marginalization over $z$.)  Moreover, since the MAF loss
function is just the cross-entropy loss with the true distribution,
this means that loss functions of competing models can be compared
directly. It would therefore be worthwhile to also try to improve
performance of the basic MAF model.

In contrast to typical applications of these deep learning tools, the
space $Y$ on which all of our models are conditioned is of much higher
dimensionality than the space $X$ that we are modeling. One way to
improve performance further may be to introduce convolutional neural
networks to pre-process the strain data $y$ and compress it to lower
dimensionality. In the future, when we extend our models to treat
longer waveforms, higher sampling rates, and additional detectors,
this will become even more important.

Going forward, it will also be important to understand better the
uncertainty associated to the neural network itself, particularly in
regions of parameter space that are not well covered by training
data. Rather than taking the maximum likelihood estimate for neural
network parameters, as we did in this work, one can model them as
random variables with some probability distribution. This distribution
can be learned through variational inference, and ultimately
marginalized over, resulting in a slightly broader posterior over
system parameters~\cite{blundell2015weight,Shen:2019vep}. In our work,
overfitting was not a problem, but such approaches to neural network
uncertainty could be useful in situations where the binary system
parameter space is not well covered due to high waveform generation
costs.

As the rate of detected gravitational-wave events grows with improved
detector sensitivity, methods for rapid parameter estimation will
become increasingly desirable. For deep learning approaches to become
viable alternatives to standard inference methods, they must be
extended to cover the full space of binary system parameters, and to
treat longer duration waveforms from multiple detectors with detector
PSDs that vary from event to event. The methods discussed in this work
bring us closer to these goals.

\begin{acknowledgments}

  We would like to thank R. Cotesta, A. Ghosh, A. Matas, and
  S. Ossokine for helpful discussions.  C.S. was supported by the
  EPSRC Centre for Doctoral Training in Data Science, funded by the UK
  Engineering and Physical Sciences Research Council (grant
  EP/L016427/1) and the University of Edinburgh.

\end{acknowledgments}

\bibliography{mybib.bib}

\end{document}